\documentclass[9pt, conference]{IEEEtran}
\IEEEoverridecommandlockouts
\usepackage{cite}
\usepackage{amsmath,amssymb,amsfonts}
\usepackage{algorithmic}
\usepackage{booktabs}
\usepackage{graphicx}
\usepackage{textcomp}
\usepackage{xcolor}
\usepackage{pifont}
\usepackage{hyperref}

\newcommand{\cmark}{\ding{51}}
\newcommand{\xmark}{\ding{55}}
\newcommand{\pmark}{\textcolor{gray}{\ding{108}}} 

\def\BibTeX{{\rm B\kern-.05em{\sc i\kern-.025em b}\kern-.08em
T\kern-.1667em\lower.7ex\hbox{E}\kern-.125emX}}

\begin{document}

\title{SpeakerCard-1M: An Evidence-Grounded Corpus for In-the-Wild Speaker Verification}

\author{
\IEEEauthorblockN{Junyi Peng$^{1}$, Old\v{r}ich Plchot$^{1}$, Xiao Song$^{2}$, Dading Chong$^{2}$, Lichun Fan$^{3}$, Hang Su$^{3}$,\\ Themos Stafylakis$^{4}$, Junjie Li$^{5}$, Kong Aik Lee$^{5}$, Shuai Wang$^{6}$, Jian Luan$^{3}$, Jan {\v{C}}ernock\'{y}$^{1}$} \IEEEauthorblockA{$^{1}$Brno University of Technology, Czechia \quad $^{2}$Peking University, China \quad $^{3}$Xiaomi, China \\ $^{4}$Athens University of Economics and Business, Greece \\ $^{5}$The Hong Kong Polytechnic University, Hong Kong \quad $^{6}$Nanjing University, China} }

\maketitle

\begin{abstract}
Modern speaker verification (SV) systems rely on speaker embeddings that are effective but difficult to interpret or query in natural language.
Most existing speech--text corpora target controllable synthesis or utterance-level captioning, offering limited speaker-level supervision for in-the-wild speaker recognition.
This paper introduces \textbf{SpeakerCard-1M}, a bilingual speaker resource for evidence-grounded SV, derived from VoxCeleb1/2 and CN-Celeb1/2, where the ``-1M'' suffix refers to the 1.78M utterance-level captions contained in the release.
We adopt a \emph{tool-first, LLM-last} approach in which ten acoustic probes produce field-level evidence, the evidence is aggregated into speaker profiles under a schema that separates relatively stable \emph{traits} from utterance-level \emph{states}, and bilingual Speaker Cards are rendered by a constrained LLM that sees only the structured fields.
The release includes 56.7k Speaker Card records over 10.2k speakers, 1.78M utterance-level captions, and speaker-ID-disjoint hard-negative triplets.
We further define two SV-oriented cross-modal protocols, bidirectional Speaker--Text Retrieval (T2S-R / S2T-R) and Attribute-Conditioned Verification (AC-Verify), and compare a dual-encoder baseline against recent audio language models under a zero-shot forced-choice setting.
Joint audio--text training costs only 0.31\% absolute EER on VoxCeleb1-O relative to the audio-only baseline.
Under a style-symmetric LLM-generated counterfactual protocol, eight recent audio language models (7B--30B+ parameters, both open- and closed-source) score 49--77\% on pitch-level AC-Verify in a 2-way forced-choice setting, compared with 88.66\% for our dual encoder.
\end{abstract}


\section{Introduction}

Modern speaker verification (SV) systems map speech into compact speaker embeddings~\cite{snyder2018x,desplanques2020ecapa, chen2022wavlm} and have driven steady progress on benchmarks such as VoxCeleb~\cite{nagrani2017voxceleb,chung2018voxceleb2}.
These embeddings are effective, but a similarity score only supports a same/different decision: it says nothing about the speaker attributes a user can search for or contradict.

This becomes a critical problem as speech systems are increasingly accessed through natural-language interfaces: a user may want to look up a speaker from a verbal description, or check whether a model still agrees with an enrollment when one attribute is changed.
General large audio language models (LALMs)~\cite{tang2024salmonn,chu2024qwen2audio,xu2025qwen25omni} are not built for this purpose, with a recent benchmark reporting 22--45\% EER on standard SV trials~\cite{ren2025allmsv}.
What is missing is speaker-level supervision tied to structured, queryable evidence, rather than generic free-form captions.

We refer to this setting as \emph{evidence-grounded SV}.
Rather than replacing speaker embeddings with text, we aim to provide an additional language-accessible evidence layer for SV models.
The layer separates relatively stable speaker \emph{traits} (e.g.\ gender, pitch band, accent) from utterance-level \emph{states} (e.g.\ emotion, channel, environment).
Trait fields are aggregated across recordings and used for identity-oriented descriptions, whereas state fields describe session-level conditions and are excluded from \texttt{identity\_only} cards.
This separation is enforced by the annotation \emph{schema} rather than inferred by the LLM: the schema is a fixed set of typed trait and state fields that the LLM must fill from probe evidence, rather than generate as free-form text.

Suitable supervision for this setting is scarce.
Speech--text corpora aimed at controllable generation~\cite{guo2023prompttts, leng2023prompttts2,shimizu2024prompttts,jin2024speechcraft} mainly describe utterance-level content or style, not speaker identity.
SV corpora such as VoxCeleb and CN-Celeb~\cite{fan2020cnceleb, li2022cnceleb2} provide speaker labels and verification trials, but no natural-language description layer.
Recent speaker--text retrieval~\cite{liu2024spkrtext} and speaker profiling~\cite{baali2025colmbo,feng2025voxprofile} efforts come closer; Vox-Profile~\cite{feng2025voxprofile}, in particular, introduces a trait-versus-state taxonomy at the benchmark level.
To our knowledge, no existing resource brings these together: bilingual in-the-wild coverage, schema-level trait--state separation, per-attribute provenance linking each trait label to its underlying probe output, and SV-oriented cross-modal protocols.

\begin{table*}[t]
\centering
\caption{Comparison with representative resources. The table summarizes released annotations and evaluation protocols rather than overall dataset quality. Columns: \textbf{Wild} (in-the-wild data),
\textbf{Spk.\ text} (speaker-level rather than utterance-level
text), \textbf{T--S Sep.} (programmatically enforced trait--state
schema), \textbf{Evidence} (field-level per-attribute provenance),
\textbf{X-SV} (cross-modal SV protocols beyond audio-only trials).
\cmark{} supported, \pmark{} partial, \xmark{} not;
\emph{italics} qualified, \textbf{bold} indicates a more explicit implementation in our release.}
\label{tab:positioning}
\small
\setlength{\tabcolsep}{3pt}
\renewcommand{\arraystretch}{1.12}
\begin{tabular}{lcccccccc}
\toprule
\textbf{Resource}
& \textbf{Goal}
& \textbf{Wild}
& \textbf{Spk. text}
& \textbf{Bilingual}
& \textbf{T--S Sep.}
& \textbf{Evidence}
& \textbf{Aud--Txt}
& \textbf{X-SV} \\
\midrule
VoxCeleb1/2 (2017/18)~\cite{nagrani2017voxceleb,chung2018voxceleb2}
& SV               & \cmark & \xmark & \xmark         & \xmark         & \xmark        & \xmark & \xmark \\
CN-Celeb1/2 (2020/22)~\cite{fan2020cnceleb,li2022cnceleb2}
& SV               & \cmark & \xmark & \emph{ZH only} & \xmark         & \xmark        & \xmark & \xmark \\
PromptTTS++ (2024)~\cite{shimizu2024prompttts}
& TTS              & \xmark & \xmark & \xmark         & \xmark         & \xmark        & \cmark & \xmark \\
SpeechCraft (2024)~\cite{jin2024speechcraft}
& TTS/caption      & \xmark & \xmark & \cmark         & \xmark         & \pmark        & \cmark & \xmark \\
ParaSpeechCaps (2025)~\cite{diwan2025scaling}
& Captioning       & \pmark & \xmark & \xmark         & \xmark         & \pmark        & \cmark & \xmark \\
Speaker-Text Retr. (2024)~\cite{liu2024spkrtext}
& Retrieval        & \cmark & \cmark & \emph{EN+JA}   & \xmark         & \pmark        & \cmark & \pmark \\
CoLMbo (2025)~\cite{baali2025colmbo}
& Profiling        & \cmark & \cmark & \xmark         & \xmark         & \pmark        & \cmark & \xmark \\
Vox-Profile (2025)~\cite{feng2025voxprofile}
& Profiling        & \cmark & \pmark & \xmark         & \emph{taxonomy}& \emph{labels} & \xmark & \xmark \\
SpeakerLM (2026)~\cite{yin2026speakerlm}
& Diariz.+recog.   & \cmark & \cmark & \cmark         & \xmark         & \xmark        & \cmark & \xmark \\
\midrule
\textbf{SpeakerCard-1M}
& Evd.-SV
& \cmark
& \cmark
& EN+ZH
& \textbf{schema}
& \textbf{field}
& \cmark
& \cmark \\
\bottomrule
\end{tabular}
\end{table*}
We address these gaps with \textbf{SpeakerCard-1M}\footnote{Corpus and protocols at \url{https://junyipeng00.github.io/SpeakerCard-1M-page}}, a bilingual speaker corpus built on VoxCeleb1/2 and CN-Celeb1/2 under a \emph{tool-first, LLM-last} principle. First, acoustic probes perceive the audio and populate the trait and state fields of the schema. The LLM is then used only to turn these structured fields into fluent Speaker Cards, and never observes the raw audio. This division keeps perception in the probes and confines the LLM to verbalization, avoiding free-form hallucination of speaker attributes. Our contributions are:
\begin{itemize}\setlength\itemsep{1pt}
  \item \textbf{SpeakerCard-1M}: a bilingual (EN+ZH) speaker corpus over 10.2k speakers, with 56.7k post-QC records, 1.78M utterance captions, field-level probe provenance, and speaker-disjoint hard-negative triplets.
  \item A \emph{tool-first, LLM-last} pipeline enforcing trait--state separation at the schema level.
    \item Two SV-oriented cross-modal protocols with dual-encoder baselines and zero-shot comparisons against eight open- and closed-source LALMs: \emph{bidirectional speaker--text retrieval} (T2S-R / S2T-R), matching a speaker to a natural-language description and back; and \emph{Attribute-Conditioned Verification} (AC-Verify), testing whether a model rejects a card whose single trait is contradicted or a hard-negative near-miss. Together they establish strong baselines for evidence-grounded SV.
  \item Benchmark analyses showing that joint audio--text training preserves conventional SV performance with modest EER cost, while zero-shot LALMs remain weak on fine acoustic counterfactuals such as pitch.
\end{itemize}

\section{Related Work and Positioning}
\label{sec:related}

Table~\ref{tab:positioning} positions SpeakerCard-1M against related resources along eight dimensions (column definitions in caption).

\noindent\textbf{Speech--text corpora for controllable generation and
captioning.}
PromptTTS~\cite{guo2023prompttts}, PromptTTS++~\cite{shimizu2024prompttts}, SpeechCraft~\cite{jin2024speechcraft} and ParaSpeechCaps~\cite{diwan2025scaling} pair speech with natural-language descriptions.
Although these resources have achieved promising results for controllable synthesis and utterance-level captioning, the descriptions operate at the utterance level and are typically collected under cleaner conditions than in-the-wild SV data.
Large-scale in-the-wild curation pipelines such as Emilia~\cite{he2025emilia} provide rich automatic annotations, but these are not organized at the speaker level.
In contrast, SpeakerCard-1M provides speaker-level supervision drawn from in-the-wild SV corpora, intended for verification rather than synthesis or captioning.

\noindent\textbf{Speaker language modeling and profiling.}
In~\cite{liu2024spkrtext}, a contrastive speaker--text retriever is trained on English and Japanese speakers, yet the text supervision is free-form and does not distinguish traits from states.
In CoLMbo~\cite{baali2025colmbo}, a speaker encoder is paired with a prompt-conditioned LLM to generate descriptive profiles, again as free-form text rather than as verification supervision.
SpeakerLM~\cite{yin2026speakerlm} unifies diarization and recognition with multimodal LLMs, but its outputs are speaker labels, not descriptive evidence.
Attribute-based explainable SV~\cite{williams2019speaker,ma2025expo} studies phonetic and demographic decision factors, but does not release an evidence-grounded supervision corpus.

The closest work is Vox-Profile~\cite{feng2025voxprofile}, which proposes a linguistically motivated taxonomy of static and dynamic speech traits as a benchmark for foundation models. SpeakerCard-1M differs in three concrete ways.
First, Vox-Profile evaluates at the utterance level (even static traits such as gender, age, and accent are predicted per utterance), whereas we aggregate evidence across recordings into speaker-level supervision, the natural unit for verification.
Second, the trait--state distinction is realized as a programmatically enforced annotation schema rather than only a conceptual taxonomy.
Third, each Speaker Card retains field-level provenance back to the underlying probe evidence, so the same taxonomy is usable as training supervision rather than only as a label set for evaluation.

\noindent\textbf{Audio language models.}
Speech-aware LLMs such as SALMONN~\cite{tang2024salmonn}, Qwen-Audio / Qwen2-Audio~\cite{chu2023qwenaudio,chu2024qwen2audio}, Audio Flamingo~\cite{goel2025audioflamingo3} and recent omni-modal LLMs~\cite{xu2025qwen25omni} broaden what speech systems can do at the interface level, but as Section~\ref{sec:results} shows, they are not yet reliable on fine-grained speaker discrimination.
SpeakerCard-1M provides a complementary resource: speaker-level supervision grounded in structured probe outputs, against which LALMs can be evaluated under controllable attribute-conditioned protocols.

\begin{figure*}[t]
\centering
\includegraphics[width=\textwidth]{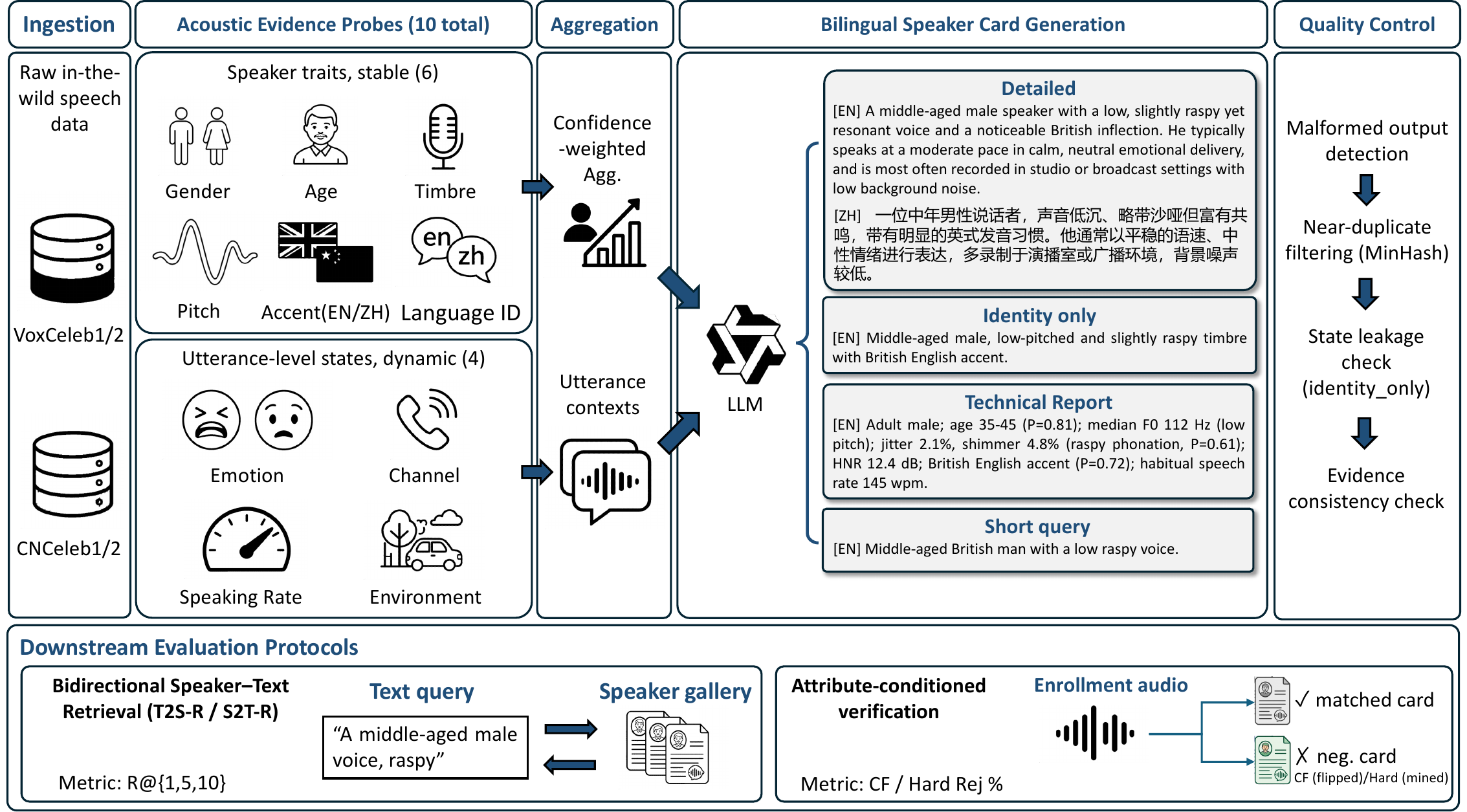}
\caption{The SpeakerCard-1M construction pipeline.
(1)~Ingestion: VoxCeleb1/2 and CN-Celeb1/2 audio is normalized
into a unified utterance manifest. (2)~Acoustic evidence probes:
ten off-the-shelf probes extract six speaker traits (gender, age,
timbre, pitch, accent, language ID) and four utterance-level states
(emotion, channel, speaking rate, environment). (3)~Aggregation:
trait fields are aggregated by confidence-weighted voting into
speaker profiles, while state fields remain per-utterance.
(4)~Bilingual Speaker Card generation: an LLM verbalizes structured
fields into four EN/ZH caption styles (\texttt{detailed},
\texttt{identity\_only}, \texttt{technical\_report},
\texttt{short\_query}) per record, without ever observing raw
audio; three paraphrase variants are sampled per (speaker,
language). (5)~Quality control: malformed-output detection,
MinHash near-duplicate filtering, state-leakage check, and
NLI-based evidence consistency check. The released cards support
two cross-modal protocols: T2S-R / S2T-R and AC-Verify.}
\label{fig:pipeline}
\end{figure*}

\section{SpeakerCard-1M Corpus Construction}
\label{sec:corpus}

\subsection{Source Data and Unified Ingestion}
\label{sec:ingest}

SpeakerCard-1M is built on top of VoxCeleb1/2~\cite{nagrani2017voxceleb,chung2018voxceleb2} and CN-Celeb1/2~\cite{fan2020cnceleb,li2022cnceleb2}.
We first consolidate the four source corpora into a single index, in which every utterance is one row with a common set of fields \texttt{\{utt\_id, speaker\_id, corpus, language\_prior, wav\_path, duration, sample\_rate\}}; we refer to this index as the utterance manifest.
We then clean the speaker pool used for aggregation: we drop corrupted files and exclude speakers with fewer than three valid utterances or fewer than 30~s of cumulative speech from speaker-level aggregation, while still keeping their utterances for utterance-level captioning.
After cleaning, the manifest contains 1.78M utterances from 10.2k speakers, totaling 4{,}069 hours.

\subsection{Probe Evidence and Trait--State Aggregation}
\label{sec:probes}

In a tool-first manner, ten automatic acoustic probes\footnote{Checkpoint URLs and revisions for neural probes, together with versions of heuristic extractors, are released with the configuration files.} extract field-level evidence covering six speaker traits and four utterance-level states (Fig.~\ref{fig:pipeline}).
Most probes are off-the-shelf models: gender and age~\cite{burkhardt2023agegender}, language ID~\cite{radford2023robust}, and speaking rate come from multilingual models; channel and environment~\cite{lavechin2023brouhaha} from language-independent acoustic extractors; timbre~\cite{teixeira2013vocal} and emotion~\cite{wang2023emotion} from English-trained voice-quality and SER models applied cross-lingually; and accent from dedicated EN and ZH~\cite{tang2021kespeech} classifiers routed by source corpus.
Pitch uses a more robust three-way consensus: we run three independent F0 systems (Praat autocorrelation via Parselmouth~\cite{boersma2001praat,jadoul2018parselmouth}, torchcrepe (CREPE pretrained, confidence threshold $0.5$)~\cite{kim2018crepe}, and RMVPE~\cite{wei23b_interspeech}), all restricted to the 75--500~Hz band, take the per-utterance median of each, and combine the three medians into one canonical F0 (median of medians) that is binned into five pitch levels.
Every probe output is stored as a structured record holding the predicted label or value, the probe's native per-utterance confidence (e.g., the predicted-class softmax probability for categorical probes), the model identifier and revision, and the source utterance ID.

Mean per-utterance probe confidence is 0.99 (language ID), 0.96 (gender), 0.75--0.82 (accent, pitch band, environment, channel), and 0.46--0.54 (age, timbre, emotion); the lower group reflects fine-grained paralinguistic difficulty and, for timbre/emotion, cross-lingual application of English-trained models.
Age and timbre are retained with hedged wording and confidence metadata; emotion is excluded from \texttt{identity\_only} cards as a state field. Confidence values are used within-field, not as calibrated probabilities across probes.

For each speaker, we combine categorical trait fields by confidence-weighted voting, and summarize continuous or ordinal fields (e.g., pitch) with robust statistics (median, MAD), binning them only when downstream verbalization requires it.
State evidence is not aggregated and remains attached to individual utterances as utterance context.

We also quantify how stable each probe is over a speaker's utterances.
For a single-label categorical field, the intra-speaker self-consistency $\sigma_{s,f}$ is the fraction of utterances whose prediction matches the aggregated label; continuous fields use a coefficient-of-variation analogue.
Unlike the per-utterance confidence above, $\sigma_{s,f}$ reflects agreement across utterances rather than accuracy or model certainty.
We flag a speaker \texttt{low\_confidence} when $\sigma_{s,\mathrm{gender}} < 0.8$, i.e., when the speaker's gender predictions are unstable across utterances; $4.52\%$ of speakers fall below this threshold, against a median of $0.997$ for the other categorical traits.

To validate probe outputs against external references, we audit a 200-speaker subset randomly sampled from the 1k-speaker AC-Verify English gallery: gender ground truth is taken from the official VoxCeleb1 metadata, and age ground truth from biographical Wikipedia entries (with the 2017 VoxCeleb1 release year as a recording-year proxy).
On this subset, the probes achieve 94.5\% gender accuracy and 74.0\% accuracy on a 4-way age-band classification (chance baseline 25\%). 
For pitch, the three-system consensus provides robustness against single-tool failures: on the audit cohort the three F0 systems agree tightly (median-F0 Pearson $r=0.95$--$0.97$, pairwise MAD $<4.5$~Hz), so the consensus F0 is internally stable, though not human-validated. The released records retain all three per-utterance F0 estimates so users can flag inter-extractor disagreement; a human-annotated pitch audit is left for future work.

\subsection{Bilingual Speaker Card Generation}
\label{sec:gen}

Speaker Cards are generated from the aggregated profiles: the LLM receives a serialized schema with trait fields, confidence and stability metadata, and (where required) a summarized utterance-state context.
The LLM never accesses raw audio and acts purely as a verbalizer: the hedging and trait/state inclusion rules of Section~\ref{sec:probes} are enforced by the prompt schema rather than left to the model.

Motivated by the heterogeneous downstream needs of retrieval and verification, we render four caption styles per record (see Section~\ref{sec:stats} for the record definition).
\texttt{detailed} combines speaker traits with a representative utterance-state context; \texttt{identity\_only} uses trait fields only and excludes states (emotion, channel, environment, speaking rate); \texttt{technical\_report} is semi-structured and reports numerical estimates and per-field confidence; and \texttt{short\_query} is a concise retrieval-oriented description.
To avoid translation artifacts, each style is rendered in English and Chinese from the same schema, by separate EN and ZH prompts rather than post-hoc translation.
Utterance-level captions are generated separately from the utterance-state fields for every retained utterance.

\textbf{Paraphrase variants.}
For each (speaker, language), Qwen2.5-72B-Instruct (temperature~$0.7$) is invoked three times under deterministic per-call seeds; a single call jointly emits all four caption styles in one JSON object, so the four fields are stylistically coherent within a variant and surface-form differences across variants reflect only sampling stochasticity.
Variants act as a noise-robustness mechanism: one variant per anchor per epoch is sampled as the positive in training; T2S-R/S2T-R uses the mean of three variants as the gallery text anchor (\S\ref{sec:eval_protocols}).

\subsection{Quality Control and Training Data Construction}
\label{sec:qc}

To control the quality of generated cards, we adopt a four-stage QC pipeline: (i) malformed-output detection; (ii) near-duplicate filtering with MinHash-LSH over character 5-grams at Jaccard threshold 0.9 (64 hashes, 16 bands), rejecting 17 cross-speaker duplicates in the current release; (iii) identity-only state-leakage rejection against a curated lexicon; and (iv) evidence consistency, scored with \texttt{MoritzLaurer/mDeBERTa-v3-base-mnli-xnli}~\cite{laurer2024multinli}, an NLI model independent of the Qwen family used for generation.
Following the standard NLI thresholding, cards below 0.6 entailment are rejected, and those in $[0.4,0.6)$ are retained but flagged.
Retained cards yield a mean entailment of 0.99 against their structured premises; rejected cards have a median entailment of 0.02 (flagged: 0.48).
Of 61.1k candidates, 56.7k (92.7\%) form the released set under tightened HF-NLI thresholding applied uniformly across all four caption fields.
The pipeline is high-recall, but does not guarantee that the retained language is free of noise.

Train, validation, and test splits are disjoint at the speaker-ID level (identity-overlap discussion in~\S\ref{sec:limitations}).
For contrastive training, we mine utterance-level triplets: easy negatives differ in coarse attributes; hard negatives share coarse traits (gender, age band, accent) but differ in finer ones (pitch band, timbre), with less restrictive fallbacks.

\subsection{Release Statistics}
\label{sec:stats}

Table~\ref{tab:corpus_stats} summarizes the release.
The post-QC card set has 56.7k records over 10.2k speakers, balanced across English and Chinese (28.3k~/~28.4k); contrastive triplets cover 189k training instances.

We use \emph{Speaker Card record} to denote one speaker--language--paraphrase item; each record contains four caption fields (\texttt{detailed}, \texttt{identity\_only}, \texttt{technical\_report}, and \texttt{short\_query}).
The 61{,}128 candidate records therefore correspond to $10{,}188\text{ speakers}\times 2\text{ languages}\times 3\text{ paraphrase variants}$, not to style-level records, and QC is applied at the record level: a record is either retained with all four caption fields or rejected as a whole.

\begin{table}[t]
\centering
\caption{SpeakerCard-1M release statistics.}
\label{tab:corpus_stats}
\small
\setlength{\tabcolsep}{4pt}
\begin{tabular}{lr}
\toprule
\textbf{Component} & \textbf{Count} \\
\midrule
Speakers (post-filter)              & 10{,}188 \\
Utterances with captions            & 1{,}783{,}791 \\
Total audio hours                   & 4{,}069.1 \\
Probe families (6 trait + 4 state)  & 10 \\
\midrule
\multicolumn{2}{l}{\emph{Card production (10.2k spk $\times$ 2 lang $\times$ 3 variants)}} \\
Candidate cards (Stage 4)           & 61{,}128 \\
Post-QC card records (Stage 5)      & 56{,}692 \\
\quad of which EN                   & 28{,}254 \\
\quad of which ZH                   & 28{,}438 \\
Caption fields per card             & 4 styles \\
Total caption fields                & 226{,}768 \\
\midrule
Speaker-disjoint training triplets  & 189{,}201 \\
\bottomrule
\end{tabular}
\end{table}

\section{Experimental Setup and Results}
\label{sec:baseline}

\subsection{Evaluation Protocols}
\label{sec:eval_protocols}

To cover conventional and cross-modal SV in a unified manner, we employ three protocols.

\textbf{Speaker Verification (SV).}
Audio-only verification is performed with cosine scoring between enrollment and test embeddings.
Equal error rate (EER) is reported on the VoxCeleb1 trial sets~\cite{nagrani2017voxceleb} (O/E/H) to measure whether joint audio--text training degrades audio-only discrimination.

\textbf{Bidirectional Speaker--Text Retrieval (T2S-R / S2T-R).}
A held-out gallery of $G$ speakers ($G{=}1{,}000$ for English on VoxCeleb; $G{=}144$ for the Chinese clean split on CN-Celeb) is constructed from the test split.
Each gallery speaker is represented by an \emph{audio anchor} (mean embedding of three deterministic enrollment utterances) and a \emph{text anchor} (mean of its three \texttt{identity\_only} paraphrase variants in the query language).
\emph{T2S-R} scores each text anchor as a query against the $G$ audio anchors; \emph{S2T-R} symmetrically scores each audio anchor against the $G$ text anchors.
A hit at rank $k$ means the target speaker's anchor is among the top $k$.
Following~\cite{liu2024spkrtext}, we report Recall@\{1, 5, 10\}.
For the Vox+CN regime (Table~\ref{tab:cross_lingual}), the test language follows the source corpus.

\textbf{Attribute-Conditioned Verification (AC-Verify).}
AC-Verify is a 2-way forced choice between a matched card and a single distractor; we report two metrics.
\emph{CF}: the matched card is the speaker's \texttt{identity\_only} text and the distractor is a counterfactual rewrite with exactly one trait overridden (gender, age range, accent, or pitch level).
Counterfactual targets are sampled uniformly from alias-normalized canonical values other than the original (\emph{e.g.}, \emph{England}~$\to$~\texttt{uk}; \emph{35--45}~$\to$~\texttt{55--65}) under a deterministic \texttt{(card\_id, trait)} seed for byte-reproducibility.
The counterfactual text is produced offline by Qwen3-32B-Instruct~\cite{yang2025qwen3} (temperature~0, schema-validated) under a minimal-edit prompt that preserves every unflipped trait phrase verbatim, removing style asymmetry between LLM-generated positives and template-rendered negatives.
Up to four trials per anchor (one per trait with valid probe evidence) are pooled across traits.
\emph{Hard}: the distractor is a speaker-ID-disjoint hard-negative card sharing coarse traits (gender, age band, accent) but differing in pitch or timbre (\S\ref{sec:qc}).
External LALMs see both cards as ``Card~A''/``Card~B'' under a single forced-choice prompt (identical across CF and Hard) with balanced A/B order; system messages, decoding, and parsing are released with the evaluation code.

We evaluate under two training settings.
\textbf{Vox-only} (Table~\ref{tab:main}) trains on VoxCeleb-derived triplets and evaluates on the VoxCeleb1 trial sets and the 1k-speaker English gallery.
\textbf{Vox+CN} (Table~\ref{tab:cross_lingual}) compares three training settings: EN-only (VoxCeleb-derived), ZH-only (CN-Celeb-derived), and Bilingual (the union of both pools). Each setting is evaluated on the English 1k-speaker VoxCeleb gallery and the Chinese 144-speaker clean CN-Celeb gallery.

\begin{table*}[t]
\centering
\caption{Main results under the Vox-only regime: SV EER (\%) on
VoxCeleb1-O/E/H; T2S-R / S2T-R as Recall@\{1,5,10\} on a
1k-speaker English gallery; AC-Verify CF and Hard (\%). ``--''
marks metrics not applicable; the Vox+CN regime is in
Table~\ref{tab:cross_lingual}.}
\label{tab:main}
\footnotesize
\setlength{\tabcolsep}{3pt}
\renewcommand{\arraystretch}{1.08}
\begin{tabular}{lccc ccc ccc cc}
\toprule
& \multicolumn{3}{c}{\textbf{SV EER}}
& \multicolumn{3}{c}{\textbf{T2S-R}}
& \multicolumn{3}{c}{\textbf{S2T-R}}
& \multicolumn{2}{c}{\textbf{AC-Verify}} \\
\cmidrule(lr){2-4}\cmidrule(lr){5-7}\cmidrule(lr){8-10}\cmidrule(lr){11-12}
\textbf{Model}
& \textbf{Vox1-O} & \textbf{Vox1-E} & \textbf{Vox1-H}
& \textbf{R@1} & \textbf{R@5} & \textbf{R@10}
& \textbf{R@1} & \textbf{R@5} & \textbf{R@10}
& \textbf{CF} & \textbf{Hard} \\
\midrule
ECAPA-TDNN~\cite{desplanques2020ecapa} &
0.80 & 0.99 & 1.87 & -- & -- & -- & -- & -- & -- & -- & -- \\
WavLM-Base SV~\cite{chen2022wavlm} &
0.84 & 0.92 & 1.75 & -- & -- & -- & -- & -- & -- & -- & -- \\
\midrule
Cascade (probe$\rightarrow$LLM card) & -- & -- & -- & 3.50 & 10.10 & 15.60 & 1.60 & 6.10 & 9.30 & -- & -- \\
\midrule
Ours (audio only)      & 0.76 & 0.79 & 1.58 & --  & --  & --   & --  & --  & --   & --   & --   \\
Ours (balanced)        & 1.07 & 0.91 & 2.07 & 3.00  & 15.30  & 24.80 & 4.60  & 16.00  & 25.50 & 93.84 & 72.53 \\
Ours (retrieval-spec.) & 1.25 & 1.07 & 2.38 & 5.10  & 16.60  & 27.50 & 5.50  & 16.90  & 27.30 & 85.45   &  65.53 \\
\bottomrule
\end{tabular}
\end{table*}

\begin{table}[t]
\centering
\caption{AC-Verify zero-shot forced-choice under the LLM-generated
counterfactual (\texttt{llm\_cf}) protocol. External models choose
between matched/contradicted cards given audio; our dual encoder
uses cosine compatibility. Trait columns: coarse demographic to
fine acoustic. CF: aggregate counterfactual rejection; Hard:
rejection of mined near-miss cards.}
\label{tab:lalm}
\footnotesize
\setlength{\tabcolsep}{2.5pt}
\renewcommand{\arraystretch}{1.08}
\resizebox{\columnwidth}{!}{%
\begin{tabular}{lrrrrrr}
\toprule
\textbf{Model} &
\textbf{Gender} & \textbf{Accent} & \textbf{Age} & \textbf{Pitch} &
\textbf{CF} & \textbf{Hard} \\
\midrule
\multicolumn{7}{l}{\emph{Open-source}} \\
Audio Flamingo~3~\cite{goel2025audioflamingo3}      & 94.59 & 71.88 & 56.06 & 55.26 & 69.45 & 50.05 \\
Qwen2-Audio-7B-Instruct~\cite{chu2024qwen2audio}    & 53.97 & 46.28 & 52.99 & 49.20 & 50.61 & 49.77 \\
Qwen3-Omni-30B-A3B-Instruct~\cite{qwenteam2025qwen3omni} & 97.76 & 95.37 & 80.93 & 69.59 & 85.91 & 55.28 \\
MiMo-Audio-7B-Instruct~\cite{xiaomi2025mimoaudio}   & 97.45 & 70.12 & 67.45 & 70.27 & 76.32 & 51.90 \\
Kimi-Audio-7B-Instruct~\cite{kimiteam2025kimiaudio} & 94.90 & 81.51 & 64.44 & 65.07 & 76.51 & 48.58 \\
\midrule
\multicolumn{7}{l}{\emph{Closed-source}} \\
Gemini~2.5 Flash~\cite{geminiteam2023gemini}        & 96.73 & 94.62 & 75.18 & 74.74 & 85.32 & 53.41 \\
Gemini~3.5 Flash~\cite{geminiteam2023gemini}        & 97.41 & 92.35 & 84.40 & 76.99 & 87.79 & 51.72 \\
GPT~audio~mini\footnotemark                         & 87.50 & 60.92 & 71.41 & 70.26 & 72.52 & 49.32 \\
\midrule
\textbf{Ours (balanced)}                           & \textbf{95.93} & \textbf{97.43} & \textbf{93.33} & \textbf{88.66} & \textbf{93.84} & \textbf{72.53} \\
\bottomrule
\end{tabular}}
\end{table}

\subsection{Baseline System}
\label{sec:baseline_system}

Our dual-encoder baseline consists of two main components: an audio tower and a text tower that map audio and text into a shared 256-dimensional space.
The audio tower is WavLM-Base~\cite{chen2022wavlm} with MHFA pooling~\cite{peng2023attention,peng2025camhfa}, initialized from a WeSpeaker~\cite{wang2023wespeaker} SV checkpoint.
The text tower is BGE-M3~\cite{chen2024bge} with attention-masked mean pooling.
To preserve the strong pretrained representations from both modalities, we freeze the encoders across all conditions, and update only the MHFA and projection heads on the audio side and the projection head on the text side.

\textbf{Cascade baseline.}
As a no-audio-tower reference, we report a cascade that performs retrieval entirely in the BGE-M3 text--text space.
The cascade reuses the same probe set, the same LLM verbalizer, and the same prompt schema as the corpus construction pipeline (\S\ref{sec:gen}): per-utterance probe outputs are rendered into fluent \texttt{identity\_only} cards on the query side, and speaker-level aggregated profiles are rendered into the same card style on the gallery side.

\textbf{Training configurations.}
\begin{itemize}
    \item \emph{Ours (audio-only)}: only the audio tower is trained
    with AAM-Softmax~\cite{deng2019arcface} ($m{=}0.2$, $s{=}32$), serving as the audio-only
    reference against which the dual-task SV degradation is measured.
    \item \emph{Ours (balanced)}: motivated by the trade-off between SV discrimination and cross-modal alignment, we train the dual encoder on
    \texttt{identity\_only} cards with three text candidates per
    anchor (positive, easy negative, hard negative), under an
    InfoNCE~\cite{oord2018representation} plus AAM objective
    $\mathcal{L} = \mathcal{L}_{\mathrm{InfoNCE}}
    + 0.2\,\mathcal{L}_{\mathrm{AAM}}$ (AAM warmup 1{,}000 steps).
    Batches are speaker-balanced at 6 speakers $\times$ 4
    utterances per GPU (global 192 audio / 576 text on 8 GPUs);
    training runs 10{,}000 steps with MUSAN~\cite{snyder2015musan} and
    RIR~\cite{ko2017rir} augmentation at $p{=}0.6$.
    \item \emph{Ours (retrieval-specialized)}: a two-stage variant that pushes retrieval recall further. We first train a retrieval-oriented base (a dual encoder trained on \texttt{technical\_report} positives), then continue from it for 3{,}000 steps with the same primary positive but adding a template-rendered positive and template-normalized negatives.
    Following~\cite{deng2019arcface,oord2018representation}, we add a hard-margin ranking loss (margin~$0.1$, weight~$0.1$,
    start step~$1{,}000$, $1{,}000$-step warmup), giving
    $\mathcal{L} = \mathcal{L}_{\mathrm{InfoNCE}}
    + 0.2\,\mathcal{L}_{\mathrm{AAM}}
    + 0.1\,\mathcal{L}_{\mathrm{hard}}$.
    The batch is reshaped to 12 speakers $\times$ 2 utterances per
    GPU with four text candidates per anchor (global 192 audio /
    768 text).
\end{itemize}

\footnotetext{\url{https://platform.openai.com/docs/models/gpt-audio-mini}}

\subsection{Results}
\label{sec:results}

\textbf{Standard SV is preserved at a modest absolute cost.}
The balanced dual-task model achieves 1.07\%/0.91\%/2.07\% EER on VoxCeleb1-O/E/H, against 0.76\%/0.79\%/1.58\% for the audio-only reference (0.31\% absolute increase on Vox1-O); the audio-only variant remains competitive with ECAPA-TDNN (0.80\%/0.99\%/1.87\%)~\cite{desplanques2020ecapa} and a WavLM-Base SV baseline (0.84\%/0.92\%/1.75\%)~\cite{chen2022wavlm}.
The retrieval-specialized variant sacrifices more SV performance (1.25\%/1.07\%/2.38\%), making the recall--SV trade-off explicit.

\textbf{Where the audio tower contributes.}
The cascade replaces the audio tower with the same probe$\rightarrow$LLM pipeline used in corpus construction (\S\ref{sec:gen}), matched in BGE-M3 text--text space; this is an architecture-matched control given BGE-M3's text-retrieval strength.
The cascade trails the dual encoder with a substantially larger gap on S2T (R@10 9.30\% vs.\ 25.50\%) than on T2S (15.60\% vs.\ 24.80\%), indicating that audio embeddings carry utterance-to-speaker aggregation information that a single-utterance probe-then-verbalize approach cannot fully recover.

\textbf{LALMs on AC-Verify.}
We compare the dual encoder with eight LALMs spanning 7B--30B+ parameters and both open- and closed-source families in Table~\ref{tab:lalm}.
Under the LLM-generated counterfactual protocol, gender is reliable on six of eight models (94--98\%), while \emph{pitch} counterfactuals remain a systematic weakness across models.
All eight LALMs score between 49.20\% and 76.99\% on pitch (mean 66.42\%), with Qwen2-Audio just below chance under 2-way forced choice; the strongest LALM on pitch is Gemini~3.5~Flash at 76.99\%.
The proposed dual encoder achieves 88.66\% on pitch, outperforming the strongest LALM by an absolute margin of 11.7\%.
Aggregated across traits, the dual encoder yields 93.84\% CF and 72.53\% Hard, against the strongest LALM at 87.79\% CF (Gemini~3.5~Flash) and 55.28\% Hard (Qwen3-Omni).
Across model scale and open/closed availability, LALMs do not yet reliably ground fine acoustic traits.

\textbf{Per-trait breakdown reveals an attribute hierarchy.}
Six of eight LALMs handle gender well (94--98\%); accent shows a clear split with Qwen3-Omni and both Gemini variants above 92\% versus 46--82\% for the rest; age sees only Qwen3-Omni (80.93\%) and Gemini~3.5~Flash (84.40\%) above 80\% with the rest at 53--75\%; pitch is the universal weak point with all eight below 77\%.
Qwen2-Audio is consistently near chance across all four traits ($46$--$54\%$), a separate failure class rather than a pitch-specific one. The pitch weakness of the remaining models is one of resolution, not blindness: rejection is near the $50\%$ chance floor for adjacent-bin flips but reaches $82$--$94\%$ three bins away, rising monotonically with bin-distance across open and closed $7$B--$30$B+ models. AudioFlamingo3 patterns with Qwen2-Audio here, near chance at every distance.

\textbf{Recall and discrimination trade off; caption style is the knob.}
The retrieval-specialized variant lifts T2S/S2T-R@10 by 2.7\%/1.8\% but AC-Verify drops on both: CF $93.84\!\rightarrow\!85.45\%$, pitch CF $88.66\!\rightarrow\!75.68\%$, Hard $72.53\!\rightarrow\!65.53\%$, indicating that recall-side tuning encourages shortcuts the CF and Hard protocols are designed to expose, with pitch most affected.
Switching only the evaluation-time text view changes retrieval but leaves AC-Verify flat (Table~\ref{tab:text_view}\,(a)): \texttt{identity\_only} matches training and gives the strongest T2S@10/S2T@10 (24.80\%/25.50\%); other views degrade accordingly while Hard varies by under 0.4\% and CF stays at 93.84\% across all four views, since both matched and counterfactual cards are scored under the same view.

\textbf{Schema enforcement protects both CF and Hard, with retrieval gains.}
Removing schema (training on \texttt{detailed} under identical hyperparameters and seed, Table~\ref{tab:text_view}\,(b)) drops CF $93.84\!\rightarrow\!86.97\%$ ($-6.87\%$), Hard $72.53\!\rightarrow\!68.03\%$ ($-4.50\%$, $-6.2\%$ rel.), and retrieval (T2S@10 $24.80\!\rightarrow\!19.30\%$, S2T@10 $25.50\!\rightarrow\!22.10\%$).
Under the stricter LLM-generated CF protocol, schema enforcement protects CF at least as much as Hard, confirming that programmatic trait--state separation contributes beyond its taxonomy role.

\textbf{Cross-lingual transfer is asymmetric.}
Under the Vox+CN regime (Table~\ref{tab:cross_lingual}), monolingual training transfers poorly ZH$\rightarrow$EN (T2S@10\,=\,7.50\%, S2T@10\,=\,6.90\%; the only direction where S2T falls below T2S, likely due to ZH text tower / EN audio mismatch) but reasonably well EN$\rightarrow$ZH (T2S@10\,=\,31.94\%, S2T@10\,=\,31.25\%), reflecting the larger EN training pool and broader EN coverage of the text tower.
Bilingual training preserves EN retrieval (T2S@10\,=\,22.40\% vs.\ 25.00\% EN-only) and substantially improves ZH (S2T@10\,=\,62.50\% vs.\ 31.25\% EN-only and 59.03\% ZH-only).
AC-Verify CF drops on unseen-language tests (EN$\rightarrow$ZH: 70.34\%; ZH$\rightarrow$EN: 71.30\%) relative to in-language CF (93.90/90.49\%), indicating that attribute-level reasoning is more language-specific than coarse cross-modal matching.

\begin{table}[t]
\centering
\caption{Text-view ablations.
(a)~Eval-time view switching on a fixed checkpoint; AC-Verify CF is
invariant (93.84\%) and omitted.
(b)~Schema-enforcement ablation: both variants share all
hyperparameters, seed, and evaluation protocol; only the training
text view differs.}
\label{tab:text_view}
\footnotesize
\setlength{\tabcolsep}{3pt}
\renewcommand{\arraystretch}{1.05}

\textit{(a) Eval-time view switching (caption style)}\\[1pt]
\begin{tabular}{lrrrrr}
\toprule
\textbf{Style} & \textbf{T2S@1} & \textbf{T2S@10} & \textbf{S2T@1} & \textbf{S2T@10} & \textbf{Hard} \\
\midrule
\texttt{detailed}         & 2.30 & 14.80 & 2.30 & 16.10 & 72.30 \\
\texttt{identity\_only}   & 3.00 & 24.80 & 4.60 & 25.50 & 72.53 \\
\texttt{technical\_report}& 0.70 &  8.60 & 2.50 & 15.30 & 72.17 \\
\texttt{short\_query}     & 2.70 & 16.80 & 3.40 & 16.40 & 72.43 \\
\bottomrule
\end{tabular}

\vspace{4pt}
\textit{(b) Training-time schema enforcement}\\[1pt]
\setlength{\tabcolsep}{4pt}
\begin{tabular}{lrrrr}
\toprule
\textbf{Training view} & \textbf{T2S@10} & \textbf{S2T@10} & \textbf{CF} & \textbf{Hard} \\
\midrule
\texttt{detailed} (no schema)          & 19.30 & 22.10 & 86.97 & 68.03 \\
\texttt{identity\_only} (schema, ours) & \textbf{24.80} & \textbf{25.50} & \textbf{93.84} & \textbf{72.53} \\
\bottomrule
\end{tabular}
\end{table}

\begin{table}[t]
\centering
\caption{Vox+CN bilingual-extension and cross-lingual transfer
under the balanced dual-task recipe. EN evaluation uses the
1k-speaker VoxCeleb gallery; ZH evaluation uses the 144-speaker
clean CN-Celeb gallery. ``Gal.'' is gallery size; EN and ZH denote
English and Chinese test subsets.}
\label{tab:cross_lingual}
\footnotesize
\setlength{\tabcolsep}{2.6pt}
\renewcommand{\arraystretch}{1.08}
\resizebox{\columnwidth}{!}{%
\begin{tabular}{llrrrrrrr}
\toprule
\textbf{Train} & \textbf{Test} & \textbf{Gal.} &
\textbf{T2S@1} & \textbf{T2S@10} & \textbf{S2T@1} & \textbf{S2T@10} & \textbf{CF} & \textbf{Hard} \\
\midrule
EN        & EN  & 1000 &  3.50 & 25.00 &  4.90 & 25.40 & 93.90 & 72.13 \\
EN        & ZH  & 144  &  7.64 & 31.94 &  2.78 & 31.25 & 70.34 & 70.37 \\
ZH        & EN  & 1000 &  1.00 &  7.50 &  0.80 &  6.90 & 71.30 & 62.20 \\
ZH        & ZH  & 144  & 13.19 & 57.64 & 10.42 & 59.03 & 90.49 & 74.77 \\
Bilingual & EN  & 1000 &  3.30 & 22.40 &  4.90 & 25.60 & 93.60 & 73.97 \\
Bilingual & ZH  & 144  & 10.42 & 53.47 &  8.33 & 62.50 & 90.67 & 74.77 \\
\bottomrule
\end{tabular}}
\end{table}

\section{Limitations and Responsible Release}
\label{sec:limitations}

\textbf{Probe-bound annotation quality.} Annotation quality is upper-bounded by the probes: coarse probes (gender, language ID, environment, channel) are reliable, while accent (especially Mandarin), timbre, and emotion remain noisier in in-the-wild conditions; the latter two are additionally applied cross-lingually.
Per-field confidence and stability metadata let downstream users weight or filter low-confidence attributes; released cards are not treated as gold-standard trait labels.
The 200-speaker biographical audit (\S\ref{sec:probes}) and three-system pitch consensus partially validate probe accuracy, but a larger human-labeled pitch audit remains future work.

\textbf{Identity-overlap and hard-negative calibration.} Splits are disjoint at the speaker-ID level of the source corpora.
We did not apply post-hoc identity-overlap deduplication using external lists such as SpeechBrain or VoxCeleb-Enhanced; near-duplicate \emph{cards} are removed by the MinHash stage but duplicate \emph{identities} across VoxCeleb speaker IDs are not.
Hard-negative card difficulty is also not calibrated against human judgments in the current release.


\textbf{Dual-use considerations.} SpeakerCard-1M inherits its source corpora's licensing terms.
We release the annotation layer (cards, probe evidence, splits, triplets) but not audio.
The \texttt{technical\_report} style produces quantitative summaries intended as a queryable artifact for interpretability research, not as forensically valid speaker characterization.
We do not endorse use of the corpus for identification, surveillance, or voice cloning, and we release no identity-linked metadata beyond what the source corpora already contain.

\section{Conclusion}
\label{sec:conclusion}

\textbf{SpeakerCard-1M} is a bilingual speaker corpus for evidence-grounded SV, built under a tool-first, LLM-last pipeline with schema-enforced trait--state separation, and paired with T2S-R/S2T-R and AC-Verify protocols.
Three observations follow: (i)~eight LALMs (7B--30B+ parameters, open and closed source) remain weak on fine acoustic contradictions, scoring 49.20--76.99\% on pitch versus 88.66\% for our dual encoder; (ii)~a probe-text cascade trails T2S retrieval and falls further behind on S2T, showing that audio embeddings carry information about utterance-to-speaker aggregation that single-utterance probe outputs followed by LLM verbalization cannot fully recover; (iii)~joint audio--text training costs only 0.31\% absolute EER on VoxCeleb1-O, while removing schema enforcement drops AC-Verify Hard by 4.5\% and CF by 6.9\%, confirming the role of trait--state separation.
We advocate reporting bidirectional retrieval together with hard-negative and attribute-conditioned tests for evidence-grounded SV.
The corpus, probe evidence, splits, triplets, and protocols will be released.

\bibliographystyle{IEEEtran}
\bibliography{mybiib}

\end{document}